\documentclass[12pt]{article}
\usepackage{sw20aip}

\input tcilatex
\QQQ{Language}{
American English
}

\begin{document}

\author{A.I.Volokitin$^{1,2}$ and B.N.J.Persson$^1$ \\
\\
$^1$Institut f\"ur Festk\"orperforschung, Forschungszentrum \\
J\"ulich, D-52425, Germany\\
$^2$Samara State Technical University, 443100 Samara,\\
Russia}
\title{Dissipative Van der Waals interaction between a small particle and a metal
surface}
\maketitle

\begin{abstract}
We use a general theory of the fluctuating electromagnetic field to
calculate the friction force acting on a small neutral particle, e.g., a
physisorbed molecule, or a nanoscale object with arbitrary dispersive and
absorptive dielectric properties, moving near a metal surface. We consider
the dependence of the electromagnetic friction on the temperature $T$, the
separation $d$, and discuss the role of screening, non-local and retardation
effects. We find that for high resistivity materials, the dissipative van
der Waals interaction can be an important mechanism of vibrational energy
relaxation of physisorbed molecules, and friction for microscopic solids.
Several controversial topics related to electromagnetic dissipative shear
stress is considered. The problem of local heating of the surface by an STM
tip is also briefly commented on.
\end{abstract}

\section{ Introduction}

There are an increasing number of investigations of the dissipative van der
Waals interaction, and the heat transfer between nanostructures, due to the
evanescent fluctuating electromagnetic field \cite{Tomassone,Pendry1,Persson
and
Zhang,Volokitin,Dedkov,Dedkov1,Pendry,Persson1,Volokitin2,Volokitin1,Dorofeev1}
in connection with scanning probe microscopy \cite{Gostman,Dorofeev,Stipe},
the quartz-crystal microbalance \cite{Watts,Dayo}, the surface force
apparatus technique \cite{Israelachvili}, the frictional drag experiments
between 2D electron system \cite{Gramila} and the scanning thermal
microscopy under ultrahigh vacuum conditions \cite{Majumdar}.

The general theory of the fluctuating electromagnetic field was developed by
Rytov \cite{Rytov} and applied for studying the conservative and dissipative
part of the van der Waals interaction \cite
{Volokitin,Dedkov,Dedkov1,Lifshitz}, the heat transfer between macroscopic
bodies and nanostructures \cite{Volokitin2,Van Hove,Levin1,Levin2} and the
frictional drag force between 2D-electron systems \cite{Volokitin1}. Despite
numerous contributions to the literature, the field of the dissipative van
der Waals interaction is very controversial, and some fundamental problems
remain unsolved. In particular, the authors of Ref. \cite{Dedkov,Dedkov1}
(and reference therein) argue that the linear (in the sliding velocity)
friction force between bodies, due to dissipative van der Waals interaction,
is non- vanishing at zero temperature, in sharp contrast to the results of
other authors \cite{Tomassone,Pendry1,Volokitin,Volokitin2}. In Appendix A
we show that the basic equation in \cite{Dedkov,Dedkov1} is incorrect, and
that a correct treatment gives a vanishing linear friction at $T=0\,K$. The
friction force acting on a particle moving near a wall originating from the
fluctuating electromagnetic field was considered recently in \cite{Dorofeev1}%
. The friction force was calculated from the electromagnetic energy losses
in the metal substrate, arising from the dipole field induced by the
fluctuating electromagnetic field from the metal. However, as it is shown in
Appendix A, such an approach is incorrect because the friction force is
determined by strong compensation between several processes involving loss
and gain of electromagnetic energy, which were not considered in \cite
{Dorofeev1}. We note also that the energy losses in the metal substrate, due
to the dipole field induced by the metal field, are much smaller than the
energy losses of the metal field in the particle. There are also processes
of energy losses of the dipole field in the metal substrate due to
spontaneous fluctuation of dipole moment which were also not considered in 
\cite{Dorofeev1}.

Another common belief is that it is necessary to go beyond second order
perturbation theory when calculating the friction force acting on a
physisorbed molecules at temperatures $k_BT<<\Delta $, where $\Delta $ is
the energy of the first excited state of the molecule relative to the ground
state \cite{Tomassone,Persson2}. This conclusion follows from the formula 
\cite{Tomassone,Volokitin2} 
\begin{eqnarray}
F_{\mathit{fric}}=\frac{3\hbar V}{2\pi d^5}\int_0^\infty \mathrm{d}\omega
\left( -\frac{\partial n(\omega )}{\partial \omega }\right) \mathrm{Im}%
\left( \frac{\varepsilon (\omega )-1}{\varepsilon (\omega )+1}\right) 
\mathrm{Im}\alpha (\omega ),  \label{mone}
\end{eqnarray}
where 
\begin{equation}
n(\omega )=\frac 1{e^{\hbar \omega /k_BT}-1}
\end{equation}
is the Bose-Einstein factor, $\varepsilon (\omega )$ is a metal dielectric
constant and $\alpha (\omega )$ is the dipole polarizability. Since for
physisorbed molecules, $\mathrm{Im}\alpha (\omega )\sim \delta (\omega
-\Delta /\hbar )$, the friction force approach zero exponentially as $%
T\rightarrow 0$. Thus, $F\propto \exp (-\Delta /k_BT)$. This situation is
similar to that of the ``vacuum'' friction between two flat surfaces \cite
{Pendry1,Volokitin}, where the frictional stress involves a product of the
imaginary part of the reflection factors for both surfaces. As a result,
``vacuum'' friction is non-vanishing only between two metal surfaces: it
vanish for metal-dielectric or dielectric-dielectric systems. Formula (\ref
{mone}) does not include screening effects which become important for
sufficiently small separations between the molecule and the metal surface.

In the present paper we use the theory of the fluctuating electromagnetic
field to calculate directly the friction force acting on a fluctuating point
dipole for arbitrary direction of motion, and taking into account screening,
retardation and non-local optic effects. A striking result we find is that,
for small separations, the screening effects become crucial, and in fact
dominate the friction force between the fluctuating point dipole and the
metal surface, because they give a contribution to the friction force which
is proportional to the absolute value of the molecule polarizability instead
of it imaginary part.

There is fundamental difference between ``vacuum'' friction for two flat
surfaces and for point dipole above a flat surface. In the former case,
scattering of electromagnetic waves by the surfaces conserves the parallel
momentum. Hence, the only possible process of momentum transfer between two
flat surfaces is the emission of electromagnetic waves by one body, and
subsequent absorption by another body. For the evanescent waves this process
gives a contribution which is proportional to product of the imaginary part
of the reflection factors for both surfaces. In the case of a particle above
flat surface, the component of the momentum parallel to the surface can
change during scattering by the particle, resulting in momentum transfer.
This process gives a contribution which is proportional to product of
imaginary part of reflection factor for metal at different value of momentum
parallel to surface and absolute value of the point dipole polarizability.
The results obtained are applied to the problem of the vibrational energy
relaxation of the physisorbed molecule, and the friction and the heat
transfer between an STM tip and the metal surface.

\section{Calculation of the fluctuating electromagnetic field}

Following Lifshitz \cite{Lifshitz}, to calculate the fluctuating
electromagnetic field we use the general theory of Rytov (see Ref. \cite
{Rytov,Levin2}). This method is based on the introduction of a fluctuating
current or polarization density in Maxwell equations (analogous to, for
example, the introduction a ``random'' force in the theory of Brownian
motion of a particle). We consider a semi-infinite metal having flat surface
which coincides with $xy$ - coordinate plane, and with the $z$ - axes
pointed along the upward normal. A point dipole located at $\mathbf{r}%
_0=(0,0,d)$, performing small amplitude vibrations with the displacement
vector $\mathbf{u}(t)=\mathbf{u}_0e^{-\mathrm{i}\omega _0t}$. To linear
order in the vibrational coordinate $\mathbf{u}(t)$, the polarization
density corresponding to the point dipole can be written in the form: 
\begin{eqnarray}
\mathbf{p}(\mathbf{r},t) &=&\mathbf{p}_0\delta (\mathbf{r}-\mathbf{r}_0)e^{-%
\mathrm{i}\omega t}+\mathbf{p}_1(\mathbf{r},\omega )e^{-\mathrm{i}(\omega
+\omega _0)t}, \\
\mathbf{p}_1(\mathbf{r},\omega ) &=&\mathbf{p}_1\delta (\mathbf{r}-\mathbf{r}%
_0)-\mathbf{p}_0\mathbf{u}_0\cdot \frac \partial {\partial \mathbf{r}}\delta
(\mathbf{r}-\mathbf{r}_0),
\end{eqnarray}
where $\mathbf{p}_0=\mathbf{p}^f+\alpha (\omega )\mathbf{E}_0,\mathbf{p}%
_1=\alpha (\omega +\omega _0)\mathbf{E}_1,\mathbf{E}(t)=\mathbf{E}_0e^{-%
\mathrm{i}\omega t}+\mathbf{E}_1e^{-\mathrm{i}(\omega +\omega _0)t}$ is an
external electric field at the position of the dipole, $\alpha (\omega )$ is
the dipole polarizability, $\mathbf{p}^f$ is a fluctuating dipole moment
with spectral density function which accordingly to fluctuation-dissipation
theorem \cite{Landau} is given by: 
\begin{eqnarray}
\left\langle p_i^fp_j^{f*}\right\rangle _\omega &=&4\pi A(\omega ,T)\mathrm{%
Im}\alpha (\omega )\delta _{ij}  \label{mthree} \\
A(\omega ,T) &=&\frac \hbar {(2\pi )^2}\left( \frac 12+n(\omega )\right) ,
\end{eqnarray}
Outside the metal the electric field is given by the sum of the electric
field from the point dipole, $\mathbf{E}_d(\mathbf{r},t)$, and the electric
field from the metal induced by point dipole, $\mathbf{E}_d^{\mathit{ind}}(%
\mathbf{r},t)$, and the electric field from the metal $\mathbf{E}^s(\mathbf{r%
},t)$ in absence of point dipole and originated from thermal and quantum
fluctuation of polarization inside the metal: 
\begin{equation}
\mathbf{E}^{\mathit{total}}(\mathbf{r},t)=\mathbf{E}_d(\mathbf{r},t)+\mathbf{%
E}_d^{\mathit{ind}}(\mathbf{r},t)+\mathbf{E}^s(\mathbf{r},t)
\end{equation}
In our consideration we will need only the last two fields. The electric
field $\mathbf{E}_d^{\mathit{ind}}(\mathbf{r},t)$ can be written in the
form: 
\begin{eqnarray}
\mathbf{E}_d^{\mathit{ind}}(\mathbf{r},t) &=&\mathbf{E}_0(\mathbf{r},\omega
)e^{-\mathrm{i}\omega t}+\mathbf{E}_1(\mathbf{r},\omega +\omega _0)e^{-%
\mathrm{i}(\omega +\omega _0)t} \\
E_{0i}(\mathbf{r},\omega ) &=&D_{ij}(\mathbf{r},\mathbf{r}_0,\omega )p_{0j}
\end{eqnarray}
\begin{equation}
E_{1i}(\mathbf{r},\omega +\omega _0)=D_{ij}(\mathbf{r},\mathbf{r}_0,\omega
+\omega _0)p_{1j}+\mathbf{u}_0\cdot \frac \partial {\partial \mathbf{r}%
^{\prime }}D_{ij}(\mathbf{r},\mathbf{r}^{\prime },\omega +\omega _0)_{%
\mathbf{r}^{\prime }=\mathbf{r}_0}p_{0j}
\end{equation}
where $\tilde D_{ik}(\mathbf{r},\mathbf{r}^{\prime },\omega )=D_{ik}^0(%
\mathbf{r},\mathbf{r}^{\prime },\omega )+D_{ik}(\mathbf{r},\mathbf{r}%
^{\prime },\omega )$ obeys the equations \cite{Abrikosov} 
\begin{equation}
\left[ \nabla _i\nabla _k-\delta _{ik}(\nabla ^2+(\omega /c)^2)\varepsilon (%
\mathbf{r})\right] \tilde D_{kj}(\mathbf{r},\mathbf{r}^{\prime },\omega
)=(4\pi \omega ^2/c^2)\delta _{ij}\delta (\mathbf{r}-\mathbf{r}^{\prime })
\label{mfive}
\end{equation}
\begin{equation}
\left[ \nabla _j^{\prime }\nabla _k^{\prime }-\delta _{jk}({\nabla ^{\prime }%
}^2+(\omega /c)^2)\varepsilon (\mathbf{r}^{\prime })\right] \tilde D_{ik}(%
\mathbf{r},\mathbf{r}^{\prime },\omega )=(4\pi \omega ^2/c^2)\delta
_{ij}\delta (\mathbf{r}-\mathbf{r}^{\prime }),  \label{msix}
\end{equation}
the function $D_{ik}^0(\mathbf{r},\mathbf{r}^{\prime },\omega )$ obeys the
inhomogeneous equations (\ref{mfive}-\ref{msix}) for free space, and $D_{ik}(%
\mathbf{r},\mathbf{r}^{\prime },\omega )$ outside the metal determines an
induced electric field and obeys the homogeneous equations (\ref{mfive}-\ref
{msix}). The solution of the equations (\ref{mfive}-\ref{msix}) is described
in detail in Appendix B. The electric field from the metal $\mathbf{E}^s(%
\mathbf{r},t)=\mathbf{E}^s(\mathbf{r},\omega )e^{-\mathrm{i}\omega t}$ is
characterized by the following spectral density function \cite
{Volokitin2,Abrikosov} 
\begin{equation}
\left\langle E_i^s(\mathbf{r})E_j^{s*}(\mathbf{r}^{\prime })\right\rangle
_\omega =4\pi A(\omega ,T)\mathrm{Im}D_{ij}(\mathbf{r},\mathbf{r}^{{\prime }%
},\omega )  \label{mseven}
\end{equation}
The electric fields $\mathbf{E}_0$ and $\mathbf{E}_1$ at the position of
point dipole can be found from the condition of self-consistency: 
\begin{eqnarray}
E_{0i} &=&D_{ii}(\mathbf{r}_0,\mathbf{r}_0,\omega )p_{0i}+E_i^s(\mathbf{r}%
_0,\omega ),  \label{mnine} \\
E_{1i} &=&D_{ii}(\mathbf{r_0},\mathbf{r}_0,\omega +\omega _0)\alpha (\omega
+\omega _0)E_{1i}+\mathbf{u}_0\cdot \frac \partial {\partial \mathbf{r}}%
\Biggl(E_i^s(\mathbf{r},\omega )+  \nonumber \\
&&D_{ij}(\mathbf{r_0},\mathbf{r},\omega +\omega _0)p_{0j}+D_{ij}(\mathbf{r},%
\mathbf{r}_0,\omega )p_{0j}\Biggr)_{\mathbf{r}=\mathbf{r}_0}.  \label{mten}
\end{eqnarray}
In (\ref{mnine}-\ref{mten}) we use that $D_{ik}(\mathbf{r},\mathbf{r}%
)=\delta _{ik}D_{ii}(\mathbf{r},\mathbf{r})$. From (\ref{mnine}-\ref{mten})
we get: 
\begin{equation}
E_{0i}=\frac{E_i^s(\mathbf{r}_0)+D_{ii}(\mathbf{r}_0,\mathbf{r}_0,\omega
)p_i^f}{1-\alpha (\omega )D_{ii}(\mathbf{r}_0,\mathbf{r}_0,\omega )},
\end{equation}
\begin{equation}
p_{0i}=\frac{p_i^f+\alpha (\omega )E_i^s(\mathbf{r}_0,\omega )}{1-\alpha
(\omega )D_{ii}(\mathbf{r}_0,\mathbf{r}_0,\omega )},  \label{meleven}
\end{equation}
\begin{equation}
E_{1i}=\mathbf{\ u}_0\cdot \frac \partial {\partial \mathbf{r}}\frac{\Biggl( %
E_i^s(\mathbf{r})+D_{ij}(\mathbf{r_0},\mathbf{r},\omega +\omega
_0)p_{0j}+D_{ij}(\mathbf{r},\mathbf{r}_0,\omega )p_{0j}\Biggr)_{\mathbf{r}=%
\mathbf{r}_0}}{1-\alpha (\omega +\omega _0)D_{ii}(\mathbf{r}_0,\mathbf{r}%
_0,\omega +\omega _0)}.  \label{mtwelve}
\end{equation}

\section{Force of friction on dipole}

The total electromagnetic force on a fluctuating dipole is determined by the
Lorentz force: 
\begin{equation}
\mathbf{F}=\int_{-\infty }^\infty \mathrm{d}\omega \int \mathrm{d}^3r\left(
\left\langle \rho \mathbf{E}\right\rangle +\frac 1c\left\langle \mathbf{j}%
\times \mathbf{B}\right\rangle \right) ,  \label{thirteen}
\end{equation}
where the integration is over the volume of the dipole vibrating in the
external electric field 
\begin{equation}
\mathbf{E}(\mathbf{r},t)=\left( \mathbf{E}_0(\mathbf{r},\omega )+\mathbf{E}%
^s(\mathbf{r},\omega )\right) e^{-\mathrm{i}\omega t}+\mathbf{E}_1(\mathbf{r}%
,\omega +\omega _0)e^{-\mathrm{i}(\omega +\omega _0)t}
\end{equation}
and magnetic induction field which can be obtained from the electric field
using Maxwell's equations 
\begin{equation}
\mathbf{B}(\mathbf{r},t)=-\mathrm{i}c\mathbf{\nabla }\times \Biggl(\left( 
\mathbf{E}_0(\mathbf{r},\omega )+\mathbf{E}^s(\mathbf{r},\omega )\right) 
\frac{e^{-\mathrm{i}\omega t}}\omega +\mathbf{E}_1(\mathbf{r},\omega +\omega
_0)\frac{e^{-\mathrm{i}(\omega +\omega _0)t}}{\omega +\omega _0}\Biggr).
\end{equation}
In (\ref{thirteen}) $\rho (\mathbf{r},t)$ and $\mathbf{j}(\mathbf{r},t)$ are
the electron and current densities of the dipole which can be expressed
through the polarization density $\mathbf{p}(\mathbf{r},t)$: 
\begin{eqnarray}
\rho (\mathbf{r},t)=-\mathbf{\nabla }\cdot \mathbf{p}(\mathbf{r},t)=-\frac
\partial {\partial x_l}\left( p_{0l}(\mathbf{r},\omega )e^{-\mathrm{i}\omega
t}+p_{1l}(\mathbf{r},\omega )e^{-\mathrm{i}(\omega +\omega _0)t}\right) ,
\label{fourteen}
\end{eqnarray}
\begin{eqnarray}
\mathbf{j}(\mathbf{r},t)=\frac \partial {\partial t}\mathbf{p}(\mathbf{r}%
,t)=-\mathit{i}\left( \omega \mathbf{p}_0(\mathbf{r},\omega )e^{-\mathrm{i}%
\omega t}+(\omega +\omega _0)\mathbf{p}_1(\mathbf{r},\omega )e^{-\mathrm{i}%
(\omega +\omega _0)t}\right) .  \label{fifteen}
\end{eqnarray}
To linear order in vibrational coordinate $\mathbf{u}(t)$ and frequency $%
\omega _0$ the total force acting on the point dipole can be written in the
form 
\begin{equation}
\mathbf{F}(t)=\mathbf{F}_{st}(\mathbf{r}_0)+\mathbf{F}_{dc}(t)+\mathbf{F}_{%
\mathit{fric}}(t).  \label{jone}
\end{equation}
Here the first term determines the conservative van der Waals force at the
position $\mathbf{r}=\mathbf{r}_0$, and the second term is the change of the
conservative van der Waals force during vibration, given by 
\begin{equation}
\mathbf{F}_{dc}(t)=\mathbf{u}(t)\cdot \frac{\mathrm{d}}{\mathrm{d}\mathbf{r}%
_0}\mathbf{F}_{st}(\mathbf{r}_0)
\end{equation}
The last term in (\ref{jone}) determines the friction force: 
\begin{equation}
\mathbf{F}_{\mathit{fric}}(t)=\mathrm{i}\omega _0\eta \mathbf{u}(t)=-\eta
\dot {\mathbf{u}}(t)
\end{equation}
Using results from Appendix C for motion parallel to the surface we obtain: 
\begin{eqnarray}
\left( \mathbf{F}_{\mathit{fric}}\right) _x &=&\frac{2\hbar V}\pi
\int_0^\infty \mathrm{d}\omega \left( -\frac{\partial n}{\partial \omega }%
\right)  \nonumber \\
&&\times \Biggl[\sum_{l=x,y,z}\Biggl(\frac{\partial ^2}{\partial x\partial
x^{\prime }}\mathrm{Im}D_{ll}(\mathbf{r},\mathbf{r}^{\prime })\mathrm{Im}%
\frac{\alpha (\omega )}{1-\alpha (\omega )D_{ll}(\mathbf{r}_0,\mathbf{r}%
_0,\omega )}\Biggr)  \nonumber \\
-2 &\mid &\alpha (\omega )\mid ^2\mathrm{Re}\Biggl(\frac 1{\left( 1-\alpha
^{*}(\omega )D_{zz}^{*}(\mathbf{r}_0,\mathbf{r}_0,\omega )\right) \left(
1-\alpha (\omega )D_{xx}(\mathbf{r}_0,\mathbf{r}_0,\omega )\right) }\Biggr) 
\nonumber \\
&&\times \Biggl(\frac \partial {\partial x}\mathrm{Im}D_{xz}(\mathbf{r},%
\mathbf{r}_0)\Biggr)^2\Biggr]_{\QATOP{\mathbf{r}=\mathbf{r}_0}{\mathbf{r}%
^{\prime }=\mathbf{r}_0}}  \label{juneone}
\end{eqnarray}
and for the motion normal to the surface we obtain 
\begin{eqnarray}
\left( \mathbf{F}_{\mathit{fric}}\right) _z &=&\frac{2\hbar V}\pi
\int_0^\infty \mathrm{d}\omega \left( -\frac{\partial n}{\partial \omega }%
\right) \sum_{l=x,y,z}\Biggl\{\frac{\partial ^2}{\partial z\partial
z^{\prime }}\Biggl[\mathrm{Im}D_{ll}(\mathbf{r},\mathbf{r}^{\prime },\omega )
\nonumber \\
&&+\mathrm{Im}\Biggl(\frac{\alpha (\omega )D_{ll}(\mathbf{r},\mathbf{r}%
_0,\omega )D_{ll}(\mathbf{r}^{\prime },\mathbf{r}_0,\omega )}{1-\alpha
(\omega )D_{ll}(\mathbf{r}_0,\mathbf{r}_0,\omega )}\Biggr) \Biggr]\mathrm{Im}%
\frac{\alpha (\omega )}{1-\alpha (\omega )D_{ll}(\mathbf{r}_0,\mathbf{r}%
_0,\omega )}  \nonumber \\
&&+\Biggl(\frac \partial {\partial z}\mathrm{Im}\Biggl( \frac{\alpha (\omega
)D_{ll}(\mathbf{r},\mathbf{r}_0,\omega )}{1-\alpha (\omega )D_{ll}(\mathbf{r}%
_0,\mathbf{r}_0,\omega )}\Biggr)\Biggr)^2\Biggr\} _{\QATOP{\mathbf{r}=%
\mathbf{r}_0}{\mathbf{r}^{\prime }=\mathbf{r}_0}}  \label{junetwo}
\end{eqnarray}

\section{Force of friction on physisorbed molecules}

The friction force acting on the moving molecule usually is written in the
form 
\begin{equation}
\mathbf{F}=-M\eta \mathbf{V,}
\end{equation}
where $M$ is the mass and $\eta $ is the coefficient of friction. For a
physisorbed molecule we can neglect retardation effects. Formally this
corresponds to limit $c\rightarrow \infty $ in the formulae for the Green's
function. In Appendix B we show that in the non-retarded limit 
\begin{eqnarray}
D_{xx}(\mathbf{r},\mathbf{r}^{\prime }) &=&\int \frac{\mathrm{d}^2q}{2\pi }%
\frac{q_x^2}qR_p(\mathbf{q},\omega )e^{\mathrm{i}\mathbf{q}(\mathbf{x}-%
\mathbf{x}^{\prime })-q(z+z^{\prime })}, \\
D_{yy}(\mathbf{r},\mathbf{r}^{\prime }) &=&\int \frac{\mathrm{d}^2q}{2\pi }%
\frac{q_y^2}qR_p(\mathbf{q},\omega )e^{\mathrm{i}\mathbf{q}(\mathbf{x}-%
\mathbf{x}^{\prime })-q(z+z^{\prime })}, \\
D_{zz}(\mathbf{r},\mathbf{r}^{\prime }) &=&\int \frac{\mathrm{d}^2q}{2\pi }%
qR_p(\mathbf{q},\omega )e^{\mathrm{i}\mathbf{q}(\mathbf{x}-\mathbf{x}%
^{\prime })-q(z+z^{\prime })}, \\
D_{xz}(\mathbf{r},\mathbf{r}^{\prime }) &=&\mathrm{i}\int \frac{\mathrm{d}^2q%
}{2\pi }q_xR_p(\mathbf{q},\omega )e^{\mathrm{i}\mathbf{q}(\mathbf{x}-\mathbf{%
x}^{\prime })-q(z+z^{\prime })},
\end{eqnarray}
where $R_p$ is the reflection factor for $p-$polarized electromagnetic
waves. For physisorbed molecules we can neglect the imaginary part of the
polarizability $\alpha (\omega )$. Then, to linear order in $\alpha
(0)z_0^{-3}$, from (\ref{juneone},\ref{junetwo}) we get 
\begin{eqnarray}
\eta _{\parallel } &=&\frac{\hbar \alpha ^2(0)}{2\pi M}\int_0^\infty \mathrm{%
d}\omega \left( -\frac{\partial n}{\partial \omega }\right) \Biggl[3\frac{%
\partial ^2}{\partial z^2}\mathrm{Im}D_{zz}(z,z_0)\mathrm{Im}D_{zz}(z_0,z_0)-
\nonumber \\
&&-2\left( \frac \partial {\partial z}\mathrm{Im}D_{zz}(z,z_0)\right) ^2%
\Biggr]_{z=z_0},  \label{iione} \\
\eta _{\perp } &=&\frac{3\hbar \alpha ^2(0)}{\pi M}\int_0^\infty \mathrm{d}%
\omega \left( -\frac{\partial n}{\partial \omega }\right) \Biggl[\frac{%
\partial ^2}{\partial z^2}\mathrm{Im}D_{zz}(z,z_0)\mathrm{Im}D_{zz}(z_0,z_0)+
\nonumber \\
&&\left( \frac \partial {\partial z}\mathrm{Im}D_{zz}(z,z_0)\right) ^2\Biggr]%
_{z=z_0}.  \label{iitwo}
\end{eqnarray}
For $z_0<\min (l,\hbar v_F/k_BT)$, where $l$ is mean free path and $v_F$ is
the Fermi velocity, the reflection factor $R_p(\mathbf{q},\omega )$ must be
calculated using non-local optic. The non-local surface contribution to $%
\mathrm{Im}R_p$ is given by \cite{Persson and Zhang} 
\begin{equation}
\left( \mathrm{Im}R_p\right) _{\mathit{surf}}=2\xi \frac \omega {\omega
_p}\frac q{k_F},  \label{rfactor}
\end{equation}
where $\xi (q)$ depends on the electron density parameter $r_s$ but
typically $\xi (0)\approx 1$. Using this expression for $\mathrm{Im}R_p$ in (%
\ref{iione}, \ref{iitwo}) gives the surface contribution: 
\begin{equation}
\eta _{\parallel \mathit{surf}}=1.9\frac{\xi ^2\hbar \alpha ^2(0)}{Mz_0^8}%
\left( \frac{k_BT}{\hbar \omega _p}\right) ^2\frac 1{(k_Fz_0)^2}
\label{iithree}
\end{equation}
and $\eta _{\perp \mathit{surf}}=8.4\eta _{\parallel \mathit{surf}}$. For
physisorbed $Xe$ on a metal one typically has $k_BT/\hbar \omega _p\sim
10^{-3}$. Using (\ref{iithree}) we get $\eta _{\parallel }\sim 10^4s^{-1}$
and $\eta _{\perp }\sim 10^5s^{-1}$ These values are a factor $\sim 10^{-4}$
smaller than estimated in \cite{Persson2}. This means that in the case of
physisorption on normal metals, the dissipative van der Waals interaction is
determined by higher order processes that are not considered in the present
theory. In the case of physisorption on high-resistivity material we can
neglect non-local effects and use a local optic expression for reflection
factor 
\begin{equation}
R_p=\frac{\varepsilon (\omega )-1}{\varepsilon (\omega )+1}
\end{equation}
In the case of constant conductivity $\sigma $ the dielectric function has
the form 
\begin{equation}
\varepsilon =1+\frac{4\pi \mathrm{i}\sigma }\omega .  \label{dfunc}
\end{equation}
As was shown in \cite{Pendry1,Pendry} the maximum dissipation is achieved
approximately at 
\begin{equation}
4\pi \sigma \approx \frac{k_BT}\hbar .  \label{heatfour}
\end{equation}
At room temperature this corresponds to $\sigma \approx 10^2(\Omega m)^{-1}$%
. Conductivities of this order are typical of semimetals such as carbon, or
of a composite. In fact only a thin coating of the right conductivity would
be required and this might arise from adsorbate layers derived from the gas
phase \cite{Pendry,Volokitin2}. Using (\ref{iione},\ref{iitwo}) and the
expression for the Green's function 
\begin{equation}
D_{zz}(z,z^{\prime })=\frac{\varepsilon (\omega )-1}{\varepsilon (\omega )+1}%
\frac 2{z+z^{\prime }}
\end{equation}
we get 
\begin{equation}
\eta _{\parallel }=\frac{3\pi }{16}\frac{\hbar \alpha ^2(0)}{Mz_0^8}\left( 
\frac{k_BT}{4\pi \hbar \sigma }\right) ^2
\end{equation}
and $\eta _{\perp }=7\eta _{\parallel }$. In the case where the temperature
is tuned to the conductivity so as to give maximum friction we get for $Xe$
physisorption $\eta _{\parallel }\sim 10^9s^{-1}$ and $\eta _{\perp }\sim
10^{10}s^{-1}$.

\section{Force of friction on an STM tip}

Another possible application of the dissipative van der Waals interaction is
scanning probe spectroscopy. In the field of scanning probe spectroscopy
(STM, AFM) it is a common approximation to model the tip by a sphere with
radius $R$. Consider distances $R<<d<<d_W\sim c\hbar /k_BT$ (at $T=300\;K$ \
we have\ \ $d_W$ $\sim 10^5\,\AA $). In this case we can consider the sphere
as a point dipole and in the formulae for the Green's function (Appendix B)
we can put $p=\mathrm{i}q$. Neglecting screening, from formulae (\ref
{juneone}, \ref{junetwo}) we get 
\begin{eqnarray}
\left( \mathbf{F}_{\mathit{fric}}\right) _x &=&\frac{\hbar V}\pi
\int_0^\infty \mathrm{d}\omega \left( -\frac{\partial n(\omega )}{\partial
\omega }\right) \int_0^\infty \mathrm{d}qq^4\mathrm{e}^{-2qd}  \nonumber \\
&&\ \ \times \left\{ 2\mathrm{Im}R_p(\omega )\mathrm{Im}\alpha (\omega
)+\left( \frac \omega {cq}\right) ^2\mathrm{Im}R_s(\omega )\mathrm{Im}\alpha
(\omega )\right\}  \label{ftip}
\end{eqnarray}
and $\left( \mathbf{F}_{\mathit{fric}}\right) _z=2\left( \mathbf{F}_{\mathit{%
fric}}\right) _x$. To simplify this expression we assume that $\mid
\varepsilon (\omega )\mid >>1$ holds for all relevant frequencies. In the
limit $d<\mid \varepsilon \mid ^{-1/2}d_W$, where $\varepsilon $ is taken at
the characteristic frequency $\omega \sim k_BT/\hbar $, the $p$- and $s$ -
wave reflection factors become 
\begin{equation}
R_p\approx \frac{\varepsilon -1}{\varepsilon +1},\;\;\;\mathrm{Im}R_p\approx 
\frac{2\mathrm{Im}\varepsilon }{\mid \varepsilon \mid ^2}.  \label{dthree}
\end{equation}
\begin{equation}
R_s=\frac{q-\sqrt{q^2-\left( \omega /c\right) ^2\varepsilon }}{q+\sqrt{%
q^2-\left( \omega /c\right) ^2\varepsilon }},\;\;\mathrm{Im}R_s\approx \frac
14\left( \frac \omega {cq}\right) ^2\mathrm{Im}\varepsilon
\end{equation}
The polarizability of the sphere is determined by 
\begin{equation}
\alpha =\frac{\varepsilon _t-1}{\varepsilon _t+2}R^3,\;\;\;\mathrm{Im}\alpha
\approx \frac{3\mathrm{Im}\varepsilon _t}{\mid \varepsilon _t\mid ^2}R^3.
\label{polarizability}
\end{equation}
We describe the sphere using the same dielectric function (\ref{dfunc}) as
the substrate: 
\[
\varepsilon _t=1+\frac{4\pi \mathrm{i}\sigma _t}\omega . 
\]
In this limit we get the $p$-wave contribution 
\begin{equation}
\left( \mathbf{F}_p\right) _x\approx 3\frac{\hbar V}{d^5}\left( \frac{k_BT}{%
4\pi \hbar }\right) ^2\sigma ^{-1}\sigma _t^{-1}R^3,  \label{fone}
\end{equation}
In the case of a good conductor at room temperature $k_BT/4\pi \hbar \sigma
\sim 10^{-5}$, for a particle with $R=10\,\AA $ and mass $M=10^{-23}kg$,
moving $d=10\,\AA $ above the surface, we get rather small coefficient of
friction $\eta \sim 0.1s^{-1}$. However if the conductivity is tuned to get
the maximum friction, when $\sigma \approx \sigma _t$ and the condition (\ref
{heatfour}) is fulfilled, we obtain $\eta \sim 10^7s^{-1}$. For $\mid
\varepsilon \mid ^{-1/2}d_W<d<d_W$ we get the $p$-wave contribution 
\begin{equation}
\left( \mathbf{F}_p\right) _x\approx 14\pi ^{-5/2}\frac{\hbar V}{d^4}\left( 
\frac{k_BT}\hbar \right) ^{5/2}\sigma ^{-1/2}\sigma _t^{-1}c^{-1}R^3.
\label{ftwo}
\end{equation}
For $d<\mid \varepsilon \mid ^{-1/2}d_W$ we get the $s-$wave contribution : 
\begin{equation}
\left( \mathbf{F}_s\right) _x\approx \frac{\pi ^3}{10}\frac{\hbar V}%
dR^3\left( \frac{k_BT}{\hbar c}\right) ^4\frac \sigma {\sigma _t},
\label{fthree}
\end{equation}
and for $\mid \varepsilon \mid ^{-1/2}d_W<d<d_W$ we get: 
\begin{equation}
\left( \mathbf{F}_s\right) _x\approx 3.54\pi ^{-3/2}\frac{\hbar V}{d^4}%
\left( \frac{k_BT}\hbar \right) ^{5/2}\sigma ^{-1/2}\sigma _t^{-1}c^{-1}R^3.
\label{ffour}
\end{equation}
From (\ref{fone}) and (\ref{fthree}) we conclude that $F_s>F_p$ for $%
0.2(cd_W/\sigma )^{1/2}<d<\,\mid \varepsilon \mid ^{-1/2}d_W$. For a normal
metal at room temperature 10$^2<d<10^3\,\AA $. For $\mid \varepsilon \mid
^{-1/2}d_W<d<d_W$, $F_s\approx F_p$. To estimate $\eta $ for $R>>d$ we can
use an approximate approach. In \cite{Volokitin2} it was shown that formula (%
\ref{ftip}) can be obtained from an exact formula for frictional stress
between two flat surfaces upon going to the limiting case of rarefied media.
We can use the opposite approach to get the frictional stress between two
flat surfaces from the friction force acting on point dipole. Neglecting
screening effects, to a linear order in the velocity $V$, the frictional
stress between two flat surfaces separated by a distance $d$, the $p$-wave
contribution is determined by a formula \cite{Volokitin,Volokitin2} 
\begin{eqnarray}
\sigma _{p\parallel }=\frac{\hbar V}{2\pi ^2}\int_0^\infty \mathrm{d}\omega
\int_0^\infty \mathrm{d}qq^3\mathrm{e}^{-2qd}\mathrm{Im}R_{1p}(\omega )%
\mathrm{Im}R_{2p}(\omega )\left( -\frac{\partial n(\omega )}{\partial \omega 
}\right) ,  \label{fflat}
\end{eqnarray}
The formula (\ref{fflat}) can be obtained from formula (\ref{ftip}) upon
integration (\ref{ftip}) over the volume of the semi- infinite body 2, and
replacing 
\begin{equation}
4\pi n\alpha \rightarrow 2R_{2p}=2\frac{\varepsilon _2-1}{\varepsilon _2+1}
\label{replacement}
\end{equation}
where $n$ is the density of the rarefied media. Using the same approach for
the sphere with $R>>d$ we get 
\begin{equation}
\eta _{\parallel }=\frac \pi 6\frac \hbar {Md^3}R\left( \frac{k_BT}{4\pi
\hbar }\right) ^2\sigma ^{-1}\sigma _t^{-1}  \label{sphere}
\end{equation}
If the conductivity is tuned to the temperature then for a sphere with $%
R\sim 10^3\;\AA ,\,d\sim 10\;\AA $ and $M\sim 10^{-17}kg$ we get $\eta \sim
10^3s^{-1}$. The same formula (\ref{sphere}) can be obtained if an STM tip
is modeled by paraboloid, $z=(x^2+y^2)/2R+d$ , and integration is performed
over the volume of paraboloid as it was described above, or over the surface
of paraboloid using the formula 
\begin{equation}
F_{p\parallel }=2\pi \int_{-\infty }^\infty \mathrm{d}\rho \rho \sigma
_{p\parallel }(L(\rho ))
\end{equation}
Here $L(\rho )$ denotes the tip-sample distance as a function of the
distance $\rho $ from the tip symmetry axis and a stress tensor $\sigma
_{p\parallel }$ is taken from (\ref{fflat}). This scheme was proposed in 
\cite{Hartmann} for the calculation of the conservative van der Waals
interaction. The error of this scheme is not larger than $5-10\%$ in
practice in an atomic force microscopy experiment, and $25\%$ in a worst
case situation \cite{Apell}.

If the conductivity is tuned to the temperature to give maximum dissipation
then for the sphere with $R=200\AA ,\,d=3\AA $ we get $\eta =F/V\sim
10^{-13}Ns/m$. This result is five order of magnitude smaller than what is
observed experimentally for the friction force acting on the AFM-tip \cite
{Gostman,Dorofeev}. Thus the dissipation of the translational energy of the
AFM-tip can not be explained by Ohmic heating originating from a fluctuating
electromagnetic field. This result is in an agreement with our earlier
conclusion \cite{Persson1}. One must consider other possible mechanism of
the friction force acting on AFM-tip which can include phononic excitation
or internal friction originating from the conservative van der Waals
interaction.

\section{Local heating of a surface by an STM tip}

It was pointed out by Pendry \cite{Pendry} that the local heating of a
surface by an STM tip can be used for local modification of a surface if the
heat transfer is sufficiently great. To investigate the power of a hot tip
to heat a surface Pendry modeled the tip as a hot sphere of the same radius $%
R$ as the tip. This is a common approximation when calculating tunneling
current and the same arguments justify its use for calculating heat
tunneling. The heat flux between an STM tip and substrate can be calculated
from the rate of work of the fluctuating electromagnetic field on the
electrons of the tip which can be expressed by the formula

\begin{equation}
S=-\frac{\mathrm{d}W}{\mathrm{d}t}=-\int \mathrm{d}^3r\left\langle \mathbf{j}%
\cdot \mathbf{E}\right\rangle
\end{equation}
where integration is over the volume of the sphere. In the case $R<<d$ we
can consider the sphere as a point dipole and obtain 
\begin{eqnarray}
S &=&\mathrm{i}\int_{-\infty }^\infty \mathrm{d}\omega \omega \left\langle
p_{0l}\left( E_{0l}^{*}(\mathbf{r}_0)+E_l^{s*}(\mathbf{r}_0)\right)
\right\rangle  \nonumber \\
\ &=&\mathrm{i}\int_{-\infty }^\infty \mathrm{d}\omega \omega \sum_{l=x,y,z}%
\frac{\left\langle \left( p_l^f+\alpha (\omega )E_l^{s*}(\mathbf{r}%
_0)\right) \left( D_{ll}^{*}(\mathbf{r}_0,\mathbf{r}_0)p_l^{f*}+E_l^{s*}%
\right) \right\rangle }{\mid 1-\alpha (\omega )(D_{ll}(\mathbf{r}_0,\mathbf{r%
}_0)\mid ^2}  \nonumber \\
\ &=&2\hbar \int_0^\infty \mathrm{d}\omega \omega \left[ n(\omega
,T_t)-n(\omega ,T)\right] \sum_{l=x,y,z}\frac{\mathrm{Im}\alpha (\omega )%
\mathrm{Im}(D_{ll}(\mathbf{r}_0,\mathbf{r}_0)}{\mid 1-\alpha (\omega )D_{ll}(%
\mathbf{r}_0,\mathbf{r}_0)\mid ^2},  \label{heatone}
\end{eqnarray}
where $T$ and $T_t$ are the temperatures of the substrate and of the tip,
respectively. Neglecting screening effects in formula (\ref{heatone}) for
distances $R<<d<<d_W$ we get \cite{Volokitin2} 
\begin{eqnarray}
S &=&2\frac \hbar \pi \int_0^\infty \mathrm{d}\omega \omega \left( n(\omega
,T_t)-n(\omega ,T)\right) \int_0^\infty \mathrm{d}qq^2\mathrm{e}^{-2qd} 
\nonumber \\
&&\ \times \left\{ 2\mathrm{Im}R_p(\omega )\mathrm{Im}\alpha (\omega
)+\left( \frac \omega {cq}\right) ^2\mathrm{Im}R_s(\omega )\mathrm{Im}\alpha
(\omega )\right\}  \label{heattwo}
\end{eqnarray}
For small separations $d<l$ we can use (\ref{rfactor}) and (\ref
{polarizability}) to get 
\begin{equation}
S=\frac{3\pi ^3\xi R^3k_B^4}{20k_Fd^4\hbar ^3\omega _p\sigma _t}\left(
T_t^4-T^4\right)
\end{equation}
Assuming $T_t=300\,K,\,d=10\,\AA ,\,R=5\,\AA ,\,\omega _p\sim
10^{16}s^{-1},\sigma _t\sim 10^{17}s^{-1}$, and that the surface is cold so
that there is no back flow of heat, we get $S\sim 10^{-15}W$. The flow of
heat will be confined to a very small area of diameter approximately the
tip- surface separation $d$, and therefore the flux per unit area $S/d^2\sim
10^3Wm^{-2}$. This should be compared to the black body heat radiation \cite
{Volokitin2,Van Hove,Levin1}: 
\begin{eqnarray}
S_{BB} &=&\frac \hbar {8\pi ^3}\int_0^\infty \mathrm{d}\omega \omega
n(\omega ,T_t)\int_{q<\omega /c}\mathrm{d}^2q\left( 1-\mid R_p(\omega )\mid
^2\right) +\left[ p\rightarrow s\right]  \nonumber \\
\ &\approx &\frac \hbar {3\pi ^2c^2}\int_0^\infty \mathrm{d}\omega \omega
^3\left( \frac \omega {4\pi \sigma }\right) ^{1/2}n(\omega ,T_t)  \nonumber
\\
\ &=&0.4\frac{k_B^4T_t^4}{\hbar ^3c^2}\left( \frac{k_BT_t}{4\pi \hbar \sigma 
}\right) ^{1/2}
\end{eqnarray}
where $[\mathrm{p}\rightarrow \mathrm{s}]$ stands for the $s$-wave
contribution obtained from the $p$-wave contribution by replacing $R_p$ with 
$R_s$. For room temperature $S_{BB}\approx 0.5\;Wm^{-2}$. Thus there is
large local enhancement of heating of the surface over that expected from
the uniform black body radiation. As pointed out in \cite{Pendry} a much
larger effect is expected for high resistivity materials. In this case the
heat flux between sphere and surface is given by \cite{Pendry,Volokitin2} 
\begin{equation}
S_p\approx \frac{2\pi ^2}5\left( \frac Rd\right) ^3\left( \frac{k_B^4}{16\pi
^2\hbar ^3\sigma \sigma _t}\right) \left( T_t^4-T^4\right) .
\label{heatthree}
\end{equation}
The heat flux (\ref{heatthree}) is maximized when the condition (\ref
{heatfour}) is fulfilled, which corresponds to conductivity $\sigma \approx
320(\Omega m)^{-1}$. In this case we get $S_p/d^2\approx 5\times
10^{10}Wm^{-2}$. Let us now assume that the tip can be modeled by a
paraboloid : $z=\left( x^2+y^2\right) /2R+d$. To get a formula for the heat
transfer between extended bodies we note that an exact formula for the heat
transfer between two semi-infinite solids ( without screening and
retardation effects) is given by the formula: 
\begin{eqnarray}
S_z=\frac \hbar {\pi ^2}\int_0^\infty \mathrm{d}\omega \omega \left(
n_1(\omega )-n_2(\omega )\right) \int_0^\infty \mathrm{d}qq\mathrm{e}^{-2qd}%
\mathrm{Im}R_{1p}(\omega )\mathrm{Im}R_{2p}(\omega )  \label{heatfive}
\end{eqnarray}
This formula can be obtained from (\ref{heattwo}) upon integration over the
volume of semi-infinite body with subsequent replacement (\ref{replacement}%
). Applying this approach to a paraboloid tip, after integration over the
volume of paraboloid and subsequent replacement of polarizability by the
reflection factor $R_{tp}$ of the tip we get 
\begin{eqnarray}
S_z &=&\frac{\hbar R}{2\pi }\int_0^\infty \mathrm{d}\omega \omega \left(
n_t(\omega )-n(\omega )\right) \int_0^\infty \mathrm{d}q\mathrm{e}^{-2qd}%
\mathrm{Im}R_p(\omega )\mathrm{Im}R_{tp}(\omega )  \nonumber \\
\ &=&\frac{\pi ^3}{15}\left( \frac Rd\right) \left( \frac{k_B^4}{16\pi
^2\hbar ^3\sigma \sigma _t}\right) \left( T_t^4-T^4\right)  \label{heatsix}
\end{eqnarray}
For the same parameters as above, and for the conductivity which gives the
maximum heat transfer (see above), we get $S_t\sim 10^{11}Wm^{-2}$.

\section{Summary and conclusion}

We have used a general theory of a fluctuating electromagnetic field to
calculate the friction force acting on a small neutral particle moving near
a metal surface taking into account screening, non-local and retardation
effects. A striking result of our study is that in the lowest order of the
perturbation theory, the dissipation of energy for sliding physisorbed
molecules is possible only due to the screening effects. The lowest order
dissipative van der Waals interaction can be an important mechanism of
vibrational energy relaxation for physisorbed molecules particularly for
high-resistivity substrates. At room temperature this implies conductivities
typical of semimetals such as carbon, or of metal-insulator composites. For
physisorbed molecules on good conductor surfaces (e.g., Gold) the
vibrational energy relaxation is dominated by higher order contributions (in
the molecule-substrate coupling), as shown in Ref. \cite{Persson2}. Another
possible application of the present theory may be to atomic force microscopy
experiments. At the optimized conditions, which at the room temperature
corresponding to the conductivity $\approx 320\,\Omega ^{-1}m^{-1}$, the
friction force acting on an AFM tip can be $F\approx \hbar VR/d^3$. In a
typical case $(R=30\,0\,\AA ,\;\,d=3\,\AA $ and $V=1\,m/s)$ this gives $%
F\approx 0.1\,pN$. This result is five order of magnitude smaller than what
was observed experimentally for the friction force acting on the AFM-tip 
\cite{Gostman,Dorofeev}. Other possible dissipation mechanisms include:
phonon and electron excitation, and the internal friction. We have discussed
several points of controversy with respect to some recent publications on
fluctuation electromagnetic dissipative shear forces. We have shown that in
the lowers order of the perturbation theory the linear (in sliding velocity)
friction force, due to dissipative van der Waals interaction, vanish at zero
temperature. Using the same approach we have calculated the heat transfer
between an STM tip and metal surface. We have found that at the same
optimized conditions as for the friction force, and at typical condition of
AFM $(d=10\,\,\AA ,\,R=10\,\,\AA ,T=300\,K\,)$, the heat flux from the tip
may be eleven order larger than the black body heat flux.

\vskip 0.5cm \textbf{Acknowledgment } A.I.V. acknowledges financial support
from DFG, Russian Foundation for Basic Research (Project No 01-02-16202).
B.N.J.P acknowledges BMBF for a grant related to the German-Israeli Project
Collaboration ``Novel Tribological Strategies from the Nano-to
Meso-Scales''. We thank J. Harris for useful comments on the text. \vskip 1cm

\appendix

\section{\thinspace}

Dedkov \textit{et. al.} have derived an expression for the friction force on
a small neutral particle moving with the velocity $V$ parallel to a flat
surface \cite{Dedkov,Dedkov1}, which is in sharp contrast to results of
other authors \cite{Tomassone,Pendry1,Persson and
Zhang,Volokitin,Persson1,Volokitin2,Volokitin1}. In particular, contrary to
the known theoretical results, the authors of Ref. \cite{Dedkov,Dedkov1}
argue that the linear (in sliding velocity) friction force between bodies,
due to a dissipative fluctuating electromagnetic field is non-vanishing at $%
T=0\;K.\,$The principal point of their theory \cite{Dedkov,Dedkov1} is the
expression for the energy dissipation rate of the fluctuating
electromagnetic field in the system ``particle- surface'': 
\begin{equation}
-\frac{\mathrm{d}W}{\mathrm{d}t}=FV=\int \left\langle \mathbf{j}^{sp}\mathbf{%
E}^{ind}\right\rangle \mathrm{d}^3r+\int \left\langle \mathbf{j}^{ind}%
\mathbf{E}^{sp}\right\rangle \mathrm{d}^3r,  \label{cone}
\end{equation}
where $\mathbf{j}^{sp}$, $\mathbf{j}^{ind}$ are the spontaneous and induced
component of the fluctuating current density in the particle; $\mathbf{E}%
^{sp}$ is the electric field outside the metal in absence of the particle; $%
\mathbf{E}^{ind}$ is the electric field from the metal induced by the
spontaneous current density $\mathbf{j}^{sp}$. However (\ref{cone}) is
incorrect for the following reason. Formula (\ref{cone}) determines the rate
of the work performed by the external electric field in the laboratory
reference frame in the volume of the particle. From energy conservation this
rate of the work is equal with opposite sign to the rate of the work of the
electric field in the volume of the metal. Since in the laboratory reference
frame the metal is at rest the latter work will be converted to the heat in
the metal volume. However the total dissipation of energy, which is
determined by the rate of the work of the force of friction, includes also
heat production in the volume of the particle, which is not included in
formula (\ref{cone}). The right relation between the rate of the work of the
force of friction and the rate of the work of the external electric field in
the volume of the particle can be obtained in the following way. The
relation between the current densities in the laboratory reference frame and
in the reference frame where the particle is at rest and the metal is moving
with velocity $-\mathbf{V}$, in the non - relativistic limit, is given by: $%
\mathbf{j}^{sp(ind)}=\mathbf{j}^{sp(ind)^{\prime }}+\rho ^{sp(ind)}\mathbf{V}
$, where $\mathbf{j}^{sp(ind)^{\prime }}$ and $\rho ^{sp(ind)}$ are the
spontaneous (induced) \ current and charge density in the moving reference
frame. Using this relation we get the relation between the rate of the work
of the electric field in the volume of the particle in the laboratory and
moving reference frame: 
\begin{eqnarray}
-\frac{\mathrm{d}W}{\mathrm{d}t} &=&\int \left\langle \mathbf{j}^{sp}\mathbf{%
E}^{ind}\right\rangle \mathrm{d}^3r+\int \left\langle \mathbf{j}^{ind}%
\mathbf{E}^{sp}\right\rangle \mathrm{d}^3r  \nonumber  \label{one} \\
\ &=&\int \left\langle \mathbf{j}^{sp^{\prime }}\mathbf{E}%
^{ind}\right\rangle \mathrm{d}^3r+\int \left\langle \mathbf{j}^{ind^{\prime
}}\mathbf{E}^{sp}\right\rangle \mathrm{d}^3r  \nonumber \\
&&\ \ \ \ \ +\mathbf{V\cdot }\left( \int \left\langle \rho ^{sp}\mathbf{E}%
^{ind}\right\rangle \right) \mathrm{d}^3r+\left( \int \left\langle \rho
^{ind}\mathbf{E}^{sp}\right\rangle \right) \mathrm{d}^3r  \nonumber \\
\ &=&-\frac{\mathrm{d}W_0}{\mathrm{d}t}+\mathbf{V\cdot F}  \label{cfour}
\end{eqnarray}
where $-\mathrm{d}W_0/\mathrm{d}t$ is the rate of the work in the moving
reference frame, which is equal to the rate of the heat production in the
volume of the particle, and $\mathbf{F}$ the total force which acts on the
particle which in the present case is equal to the friction force. We note
that the electric field in the non-relativistic limit is the same in both
reference frames. Thus, contrary to the point of view of Dedkov \textit{et.
al. }, the rate of work is determined by Eq. (\ref{cfour}), instead of Eq. (%
\ref{cone}). For a moving point dipole the spontaneous current density $%
\mathbf{j}^{\mathrm{sp}}(\mathbf{r},t)$ is given by 
\begin{eqnarray}
\mathbf{j}^{\mathrm{sp}}(\mathbf{r},t)=-\mathrm{i}\delta (z-z_0)\mathbf{p}%
^f\int \frac{\mathrm{d}^2q}{(2\pi )^2}\left( \omega +q_xV\right) e^{\mathrm{i%
}\mathbf{q}\cdot \mathbf{r}-\mathrm{i}(\omega +q_xV)t}  \label{cfive}
\end{eqnarray}
and without screening effects, as assumed in \cite{Dedkov,Dedkov1}, the
induced electric field is given by 
\begin{equation}
E_l^{\mathrm{ind}}(\mathbf{r},t)=\int \frac{\mathrm{d}^2q}{(2\pi )^2}D_{lj}(%
\mathbf{q},\omega +q_xV,z,z_0)p_j^fe^{\mathrm{i}\mathbf{q}\cdot \mathbf{r}-%
\mathrm{i}(\omega +q_xV)t}  \label{csix}
\end{equation}
Using (\ref{cfive}, \ref{csix}) we get 
\begin{eqnarray}
\int \left\langle \mathbf{j}^{sp}\mathbf{E}^{ind*}\right\rangle \mathrm{d}%
^3r &=&-8\int_0^\infty \mathrm{d}\omega \int \mathrm{d}^2qqe^{-2qz_0}A(%
\omega ,T)\times  \nonumber \\
&&\ (\omega +q_xV)\mathrm{Im}R_p(\omega +q_xV)\mathrm{Im}\alpha (\omega )
\label{cseven}
\end{eqnarray}
where we have used that in the non-retarded limit (see Appendix A): 
\begin{eqnarray}
\sum_{l=x,y,z}D_{ll}(\mathbf{q},\omega ,z,z_0)=2D_{qq}(\mathbf{q},\omega
,z,z_0)=4\pi qe^{-q(z+z_0)}R_p(\omega )  \label{ceight}
\end{eqnarray}
After the Fourier decomposition of the fluctuating electric field from the
metal 
\begin{equation}
\mathbf{E}^s(\mathbf{r},t)=\int \frac{\mathrm{d}^2q}{(2\pi )^2}\mathbf{E}^s(%
\mathbf{q},\omega ,z)e^{\mathrm{i}(\mathbf{q}\cdot \mathbf{r}-\omega t)},
\label{cten}
\end{equation}
the induced current density in the particle is given by 
\begin{eqnarray}
\mathbf{j}^{\mathrm{ind}}(\mathbf{r},t) &=&-\mathrm{i}\delta (z-z_0)\int 
\frac{\mathrm{d}^2q}{(2\pi )^2}\int \frac{\mathrm{d}^2q^{\prime }}{(2\pi )^2}%
\left[ \omega -V(q_x-q_x^{\prime })\right] \times  \nonumber \\
&&\ \alpha (\omega -q_xV)\mathbf{E}^s(\mathbf{q},\omega ,z_0)e^{\mathrm{i}%
\mathbf{q}^{\prime }\cdot \mathbf{r}-\mathrm{i}[\omega -V(q_x-q_x^{\prime
})t]}  \label{cnine}
\end{eqnarray}
From (\ref{cten},\ref{cnine}) the rate of work performed by the fluctuating
electric field from the metal on the induced current in the particle, is
given by 
\begin{equation}
\int \left\langle \mathbf{j}^{ind}\mathbf{E}^{s*}\right\rangle \mathrm{d}%
^3r=8\int_0^\infty \mathrm{d}\omega \int \mathrm{d}^2qqe^{-2qz_0}A(\omega
,T)\omega \mathrm{Im}R_p(\omega )\mathrm{Im}\alpha (\omega -q_xV)
\label{celeven}
\end{equation}
For the total rate of work performed by the electric field in the volume of
the moving particle we get 
\begin{eqnarray}
-\frac{\mathrm{d}W}{\mathrm{d}t} &=&8\int_0^\infty \mathrm{d}\omega \int 
\mathrm{d}^2qqe^{-2qz_0}A(\omega ,T)\biggl[\omega \mathrm{Im}R(\omega )%
\mathrm{Im}\alpha (\omega -q_xV)-  \nonumber \\
&&\ (\omega +q_xV)\mathrm{Im}R_p(\omega +q_xV)\mathrm{Im}\alpha (\omega )%
\biggr]  \label{dedtwelve}
\end{eqnarray}
Equation (\ref{dedtwelve}) is very similar to one obtained in \cite
{Dedkov,Dedkov1}. However, the rate of work performed by the friction force
is not determined by (\ref{cfive}) but is given by equation (\ref{cfour}).
The rate of work performed by the electric field in the volume of the
particle in the reference frame where the particle is at rest and the metal
is moving with velocity $-V$ can be found in the similar way as in the
laboratory system, and is given by 
\[
-\frac{\mathrm{d}W_0}{\mathrm{d}t}=8\int_0^\infty \mathrm{d}\omega \int 
\mathrm{d}^2qqe^{-2qz_0}A(\omega ,T)\label{ctwelve} 
\]
\begin{equation}
\times \biggl[\left( \omega -q_xV\right) \mathrm{Im}R_p(\omega )\mathrm{Im}%
\alpha (\omega -q_xV)-\omega \mathrm{Im}R(\omega +q_xV)\mathrm{Im}\alpha
(\omega )\biggr]  \label{ctwelve}
\end{equation}
From (\ref{cfour}, \ref{dedtwelve}, \ref{ctwelve}) we get the friction force 
\begin{eqnarray}
F &=&8\int_0^\infty \mathrm{d}\omega \int \mathrm{d}^2qqq_xe^{-2qz_0} 
\nonumber \\
&&\ \times A(\omega ,T)\left[ \mathrm{Im}R_p(\omega )\mathrm{Im}\alpha
(\omega -q_xV)-\mathrm{Im}R_p(\omega +q_xV)\mathrm{Im}\alpha (\omega )\right]
\nonumber \\
\ &=&\frac{2\hbar }{\pi ^2}\int_{-\infty }^\infty \mathrm{d}q_y\int_0^\infty 
\mathrm{d}q_xq_xqe^{-2qz_0}\Biggl\{ \int_0^\infty \mathrm{d}\omega \left[
n(\omega +q_xV)-n(\omega )\right]  \nonumber \\
&&\ \times \left[ \mathrm{Im}R_p(\omega )\mathrm{Im}\alpha (\omega +q_xV)+%
\mathrm{Im}R_p(\omega +q_xV)\mathrm{Im}\alpha (\omega )\right]  \nonumber \\
&&\ +\int_0^{q_xV}\mathrm{d}\omega \left( \frac 12+n(\omega )\right) 
\nonumber \\
&&\ \times \left[ \mathrm{Im}R_p(\omega )\mathrm{Im}\alpha (\omega -q_xV)+%
\mathrm{Im}R_p(\omega -q_xV)\mathrm{Im}\alpha (\omega )\right] \Bigg\}
\label{cthirteen}
\end{eqnarray}
The formula is in complete agreement with the results in \cite
{Tomassone,Pendry1,Volokitin,Volokitin2}. To linear order in the sliding
velocity $V$ we get from (\ref{cthirteen}) 
\begin{equation}
F=\frac{2\hbar V}\pi \int_0^\infty \mathrm{d}\omega \frac{\partial n(\omega )%
}{\partial \omega }\int_0^\infty \mathrm{d}qq^4e^{-2qd}\mathrm{Im}%
R_p(q,\omega )\mathrm{Im}\alpha (\omega )  \label{dipole}
\end{equation}

\section{\thinspace}

Since equations (\ref{mfive}-\ref{msix}) are translational invariant in the $%
\mathbf{x}=(x,y)$ plane, the Green's function $\tilde D_{ij}(\mathbf{r},%
\mathbf{r}^{{\prime }})$ can be represented by the Fourier integral: 
\begin{equation}
\tilde D_{ij}(\mathbf{r},\mathbf{r}^{\prime })=\int \frac{\mathrm{d}^2q}{%
(2\pi )^2}e^{\mathrm{i}\mathbf{q}\cdot (\mathbf{x}-\mathbf{x}^{\prime
})}\tilde D_{ij}(z,z^{\prime },\mathbf{q},\omega ),
\end{equation}
After the Fourier transformation it is convenient to choose the coordinate
axes in the $xy$- plane along the vectors $\mathbf{q}$ and $\mathbf{n}=\hat
z\times \mathbf{q}$. Thus (\ref{mfive}) and (\ref{msix}) become 
\begin{eqnarray}
\left( p^2+\frac{\partial ^2}{\partial z^2}\right) \tilde D_{nn}(z,z^{\prime
}) &=&-\frac{4\pi \omega ^2}{c^2}\delta (z-z^{\prime })  \label{mmone} \\
\left( \left( \frac \omega c\right) ^2+\frac{\partial ^2}{\partial z^2}%
\right) \tilde D_{qq}(z,z^{\prime })-\mathrm{i}q\frac \partial {\partial
z}\tilde D_{zq}(z,z^{\prime }) &=&-\frac{4\pi \omega ^2}{c^2}\delta
(z-z^{\prime })  \label{mmtwo} \\
p^2\tilde D_{zq}(z,z^{\prime })-\mathrm{i}q\frac \partial {\partial z}\tilde
D_{qq}(z,z^{\prime }) &=&0  \label{mmfour} \\
p^2\tilde D_{zz}(z,z^{\prime })-\mathrm{i}q\frac \partial {\partial z}\tilde
D_{qz}(z,z^{\prime }) &=&-\frac{4\pi \omega ^2}{c^2}\delta (z-z^{\prime })
\label{mmthree} \\
p^2\tilde D_{qz}(z,z^{\prime })+\mathrm{i}q\frac \partial {\partial
z^{\prime }}\tilde D_{qq}(z,z^{\prime }) &=&0  \label{mmfive}
\end{eqnarray}
\begin{equation}
p=\sqrt{\left( \frac \omega c\right) ^2-q^2}
\end{equation}
Since the equations for $\tilde D_{qn}$ and $\tilde D_{zn}$ are homogeneous,
these components of the Green function must vanish. Thus solving the system
of equations (\ref{mmone}) - (\ref{mmfive}) is equivalent to solving
altogether two equations: equation (\ref{mmone}) for $\tilde D_{nn}$, and
the equation for $\tilde D_{qq}$ which follows from equations (\ref{mmtwo})
and (\ref{mmfour}): 
\begin{equation}
\left( p^2+\frac{\partial ^2}{\partial z^2}\right) \tilde D_{qq}(z,z^{\prime
})=-4\pi p^2\delta (z-z^{\prime }).  \label{mmsix}
\end{equation}
$\tilde D_{qz}$, $\tilde D_{zq}$ and $\tilde D_{zz}$ for $z\not =z^{\prime }$
can be obtained as 
\begin{equation}
\tilde D_{qz}=-\frac{\mathrm{i}q}{p^2}\frac \partial {\partial z^{\prime
}}\tilde D_{qq};\;\tilde D_{zq}=\frac{\mathrm{i}q}{p^2}\frac \partial
{\partial z^{\prime }}\tilde D_{qq};\;\tilde D_{zz}=\frac{q^2}{p^4}\frac{%
\partial ^2}{\partial z\partial z^{\prime }}\tilde D_{qq}
\end{equation}
In the vacuum region, $z>0$, the solution of (\ref{mmone}) has the form 
\begin{equation}
\tilde D_{nn}(z,z^{\prime })=\frac{2\pi \mathit{i}\omega ^2}{pc^2}e^{\mathrm{%
i}p\mid z-z^{\prime }\mid }+v_ne^{\mathrm{i}pz}
\end{equation}
At the boundary $z=0$ , the amplitude of the reflected wave is equal to the
amplitude of the incident wave times the corresponding reflection factor.
The Green's function $\tilde D_{nn}$ is associated with the $s$-wave, and
satisfies the boundary condition 
\begin{equation}
v_n=R_s\frac{2\pi \mathit{i}\omega ^2}{pc^2}e^{\mathrm{i}pz^{\prime }}\qquad %
\mbox{for}\qquad z=0
\end{equation}
Thus 
\begin{equation}
\tilde D_{nn}(z,z^{\prime })=\frac{2\pi \mathit{i}\omega ^2}{pc^2}\Bigg\{e^{%
\mathrm{i}p\mid z-z^{\prime }\mid }+R_se^{\mathrm{i}p(z+z^{\prime })}\Bigg\}
\label{mmseven}
\end{equation}
In (\ref{mmseven}) the first and the second terms correspond to the Green's
function $D_{nn}^0$ and $D_{nn}$, respectively. Equation (\ref{mmsix}) for $%
\tilde D_{qq}$ is similar to equation (\ref{mmone}) for $\tilde D_{nn}$, and
taking into account that $\tilde D_{qq}$ is associated with $p$- wave it can
be obtained directly from (\ref{mmseven}) by replacing $R_s\rightarrow -R_p$%
: 
\begin{equation}
\tilde D_{qq}=\left( \frac{pc}\omega \right) ^2\tilde D_{nn}\left[
R_s\rightarrow -R_p\right]  \label{twentyeight}
\end{equation}
We note that in our approach the calculation of the reflection factors for $%
s $- and $p$- waves is considered as a separated problem, which , if
necessary, can be solved by taking into account non-local effects. For the
local optic case the reflection factors are determined by the well known
Fresnel formulae: 
\begin{equation}
R_p=\frac{\varepsilon p-s}{\varepsilon p+s},\,\,\;\;\;\;\;R_s=\frac{p-s}{p+s}%
,  \label{mmeight}
\end{equation}
where $\varepsilon $ is the complex dielectric constant for metal and 
\begin{equation}
s=\sqrt{\frac{\omega ^2}{c^2}\varepsilon -q^2}.  \label{mmnine}
\end{equation}
The relations between Green's functions in the $xyz$ and $qnz$ coordinate
systems are given by: 
\begin{eqnarray}
D_{xx} &=&\frac{q_x^2}{q^2}D_{qq}+\frac{q_y^2}{q^2}D_{nn}, \\
D_{xy} &=&D_{yx}=\frac{q_xq_y}{q^2}\left( D_{qq}-D_{nn}\right) , \\
D_{xz} &=&\frac{q_x}qD_{qz},
\end{eqnarray}
and so on.

\section{\thinspace}

The forces from the electric and magnetic induction fields in (\ref{thirteen}%
), to linear order in the vibrational coordinate $\mathbf{u}(t)$, can be
written in the form: 
\[
\int_{-\infty }^\infty \mathrm{d}\omega \int \mathrm{d}^3r\left\langle \rho 
\mathbf{E}^{*}\right\rangle =\int_{-\infty }^\infty \mathrm{d}\omega \int 
\mathrm{d}^3r\left\langle p_l(\mathbf{r},t)\frac \partial {\partial x_l}%
\mathbf{E}^{*}(\mathbf{r},t)\right\rangle 
\]
\[
=\int_{-\infty }^\infty \mathrm{d}\omega \frac \partial {\partial x_l}\Biggl(%
\left\langle p_{0l}\left( 1+\mathbf{u}_0\cdot \frac \partial {\partial 
\mathbf{r}}\right) \left( \mathbf{E}_0^{*}(\mathbf{r},\omega )+\mathbf{E}%
^{s*}(\mathbf{r},\omega )\right) \right\rangle 
\]
\begin{equation}
+\left[ \left\langle p_{0l}\mathbf{E}_1^{*}(\mathbf{r},\omega -\omega
_0)\right\rangle +\left\langle p_{1l}\left( \mathbf{E}_0^{*}(\mathbf{r}%
,\omega )+\mathbf{E}^{s*}(\mathbf{r},\omega )\right) \right\rangle \right]
e^{-\mathrm{i}\omega _0t}\Biggr)_{\mathbf{r}=\mathbf{r}_0},  \label{bone}
\end{equation}
\[
\frac 1c\int_{-\infty }^\infty \mathrm{d}\omega \int \mathrm{d}%
^3r\left\langle \mathbf{j}\times \mathbf{B}^{*}\right\rangle =\int_{-\infty
}^\infty \mathrm{d}\omega \Biggl\{\left\langle \mathbf{p}_0\times \mathbf{%
\nabla }\times \left( \mathbf{E}_0^{*}(\mathbf{r})+\mathbf{E}^{s*}(\mathbf{r}%
)\right) \right\rangle 
\]
\[
\Biggl[ \frac \omega {\omega -\omega _0}\Biggl( \mathbf{\nabla }\left\langle 
\mathbf{p}_0\cdot \mathbf{E}_1^{*}(\mathbf{r},\omega -\omega
_0)\right\rangle -\frac \partial {\partial x_l}\left\langle p_{0l}\mathbf{E}%
_1^{*}(\mathbf{r},\omega -\omega _0)\right\rangle \Biggr)
\]
\[
+\frac{\omega +\omega _0}\omega \Biggl(\mathbf{\nabla }\left\langle \left( 
\mathbf{p}_1+\mathbf{p}_0\mathbf{\ u}_0\cdot \frac \partial {\partial 
\mathbf{r}}\right) \cdot \left( \mathbf{E}_0^{*}(\mathbf{r})+\mathbf{E}^{s*}(%
\mathbf{r})\right) \right\rangle 
\]
\begin{equation}
-\frac \partial {\partial x_l}\left\langle \left( p_{1l}+p_{0l}\mathbf{u}%
_0\cdot \frac \partial {\partial \mathbf{r}}\right) \left( \mathbf{E}_0^{*}(%
\mathbf{r})+\mathbf{E}^{s*}(\mathbf{r})\right) \rangle \right) \Biggr) %
\Biggr]e^{-\mathrm{i}\omega _0t}\Biggr \}_{\mathbf{r}=\mathbf{r}_0}.
\label{btwo}
\end{equation}

From (\ref{bone}-\ref{btwo}) it follows that the friction force is
determined by the formula 
\begin{eqnarray}
\mathbf{F}_{\mathit{fric}} &=&\frac 12\omega _0\int_{-\infty }^\infty 
\mathrm{d}\omega \Biggl[\mathbf{\nabla }\frac \partial {\partial \omega _0}%
\Biggl(\frac \omega {\omega -\omega _0}\left\langle \mathbf{p}_0\cdot 
\mathbf{E}_1^{*}(\mathbf{r},\omega -\omega _0)\right\rangle   \nonumber
\label{bthree} \\
&&+\frac{\omega +\omega _0}\omega \left\langle \mathbf{p}_1\cdot \left( 
\mathbf{E}_0^{*}(\mathbf{r})+\mathbf{E}^{s*}(\mathbf{r})\right)
\right\rangle \Biggr)_{\omega _0=0}-c.c.\Biggr]_{\mathbf{r}=\mathbf{r}_0}
\label{bthree}
\end{eqnarray}
The calculation of (\ref{bthree}) is performed using the formulae 
\[
\Biggl(\frac \partial {\partial \omega _0}\frac{E_{1l}^{*}(\mathbf{r},\omega
-\omega _0)}{\omega -\omega _0}\Biggr)_{\omega _0=0}
\]
\[
=-\mathbf{u}_0\cdot \frac \partial {\partial \mathbf{r}^{\prime }}\Biggl[%
\frac \partial {\partial \omega }\Biggl(\frac{\alpha ^{*}(\omega )D_{lk}^{*}(%
\mathbf{r},\mathbf{r}_0,\omega )}{\omega \left( 1-\alpha ^{*}(\omega
)D_{kk}^{*}(\mathbf{r}_0,\mathbf{r}_0,\omega )\right) }\Biggr)
\]
\begin{eqnarray}
&&\times \left( E_k^{s*}(\mathbf{r}^{\prime },\omega )+D_{kk}^{*}(\mathbf{r}%
^{\prime },\mathbf{r}^{\prime },\omega )p_{0k}\right) +\frac{\alpha
^{*}(\omega )D_{lk}^{*}(\mathbf{r},\mathbf{r}_0,\omega )}{\omega \left(
1-\alpha ^{*}(\omega )D_{kk}^{*}(\mathbf{r}_0,\mathbf{r}_0,\omega )\right) }
\nonumber \\
&&\times \frac \partial {\partial \omega }\Biggl(D_{ks}^{*}(\mathbf{r}_0,%
\mathbf{r}^{\prime },\omega )\Biggr)p_{0s}^{*}+\frac \partial {\partial
\omega }\Biggl(\frac{D_{lk}^{*}(\mathbf{r},\mathbf{r}^{\prime },\omega )}%
\omega \Biggr)p_{0k}^{*}\Biggr]_{\mathbf{r}^{\prime }=\mathbf{r}_0},
\label{bfour}
\end{eqnarray}
\[
\Biggl(\frac \partial {\partial \omega _0}(\omega +\omega _0)p_{1l}\Biggr) %
_{\omega _0=0}=\Biggl(\frac \partial {\partial \omega _0}(\omega +\omega
_0)\alpha (\omega +\omega _0)E_{1l}\Biggr)_{\omega _0=0}
\]
\begin{eqnarray}
&=&\mathbf{u}_0\cdot \frac \partial {\partial \mathbf{r}^{\prime }}\Biggl[%
\frac \partial {\partial \omega }\Biggl(\frac{\omega \alpha (\omega )}{%
1-\alpha (\omega )D_{ll}(\mathbf{r}_0,\mathbf{r}_0,\omega )}\Biggr) \left(
E_l^s(\mathbf{r}^{\prime },\omega )+D_{ll}(\mathbf{r}^{\prime },\mathbf{r}%
^{\prime },\omega )p_{0l}\right) +  \nonumber \\
&&\frac{\omega \alpha (\omega )}{1-\alpha (\omega )D_{ll}(\mathbf{r}_0,%
\mathbf{r}_0,\omega )}\frac \partial {\partial \omega }\Biggl(D_{lk}(\mathbf{%
r}_0,\mathbf{r}^{\prime },\omega )\Biggr)p_{0k}\Biggr]_{\mathbf{r}^{\prime }=%
\mathbf{r}_0}.  \label{bfive}
\end{eqnarray}
Using the following expression for the spectral density functions 
\begin{equation}
\left\langle p_{0l}p_{0k}\right\rangle =4\pi A(\omega ,T)\delta _{lk}\mathrm{%
Im}\frac{\alpha (\omega )}{1-\alpha (\omega )D_{ll}(\mathbf{r}_0,\mathbf{r}%
_0,\omega )},
\end{equation}
\begin{equation}
\left\langle p_{0l}E_k^{s*}(\mathbf{r}^{\prime },\omega )\right\rangle =4\pi
A(\omega ,T)\frac{\alpha (\omega )}{1-\alpha (\omega )D_{ll}(\mathbf{r}_0,%
\mathbf{r}_0,\omega )}\mathrm{Im}D_{lk}(\mathbf{r}_0,\mathbf{r}^{\prime
},\omega ),
\end{equation}
\begin{eqnarray}
&&\left\langle \left( E_l^s(\mathbf{r}^{\prime },\omega )+D_{ll}(\mathbf{r}%
^{\prime },\mathbf{r}^{\prime },\omega )p_{0l}\right) \left( E_l^{s*}(%
\mathbf{r},\omega )+D_{lk}^{*}(\mathbf{r},\mathbf{r}_0,\omega
)p_{0k}^{*}\right) \right\rangle   \nonumber \\
&=&4\pi A(\omega ,T)\Biggl[\mathrm{Im}D_{ll}(\mathbf{r},\mathbf{r}^{\prime
},\omega )+\frac{\alpha ^{*}(\omega )}{1-\alpha ^{*}(\omega )D_{kk}^{*}(%
\mathbf{r}_0,\mathbf{r}_0,\omega )}D_{lk}^{*}(\mathbf{r},\mathbf{r}_0,\omega
)  \nonumber \\
&&\times \mathrm{Im}D_{lk}(\mathbf{r}^{\prime },\mathbf{r}_0,\omega )+D_{ll}(%
\mathbf{r}^{\prime },\mathbf{r}^{\prime },\omega )\mathrm{Im}\Biggl(\frac{%
\alpha (\omega )D_{ll}(\mathbf{r},\mathbf{r}_0,\omega )}{1-\alpha (\omega
)D_{ll}(\mathbf{r}_0,\mathbf{r}_0,\omega )}\Biggr)\Biggr],
\end{eqnarray}
and (\ref{bfour}), (\ref{bfive}) for the vibration along the $x$ - axis, (%
\ref{bthree}) is reduced to 
\begin{eqnarray}
\left( \mathbf{F}_{\mathit{fric}}\right) _x &=&u_04\pi \omega
_0\int_{-\infty }^\infty \mathrm{d}\omega A(\omega ,T)  \nonumber \\
&&\times \frac \partial {\partial \omega }\Biggl[ \sum_{l=x,y,z}\Biggl(\frac{%
\partial ^2}{\partial x\partial x^{\prime }}\mathrm{Im}D_{ll}(\mathbf{r},%
\mathbf{r}^{\prime })\mathrm{Im}\frac{\alpha (\omega )}{1-\alpha (\omega
)D_{ll}(\mathbf{r}_0,\mathbf{r}_0,\omega )}\Biggr)  \nonumber \\
-2 &\mid &\alpha (\omega )\mid ^2\mathrm{Re}\Biggl(\frac 1{\left( 1-\alpha
(\omega )^{*}D_{zz}(\mathbf{r}_0,\mathbf{r}_0,\omega )^{*}\right) \left(
1-\alpha (\omega )^{*}D_{xx}(\mathbf{r}_0,\mathbf{r}_0,\omega )\right) }%
\Biggr)  \nonumber \\
&&\times \Biggl(\frac \partial {\partial x}\mathrm{Im}D_{xz}(\mathbf{r},%
\mathbf{r}_0)\Biggr)^2\Biggr] _{\QATOP{\mathbf{r}=\mathbf{r}_0}{\mathbf{r}%
^{\prime }=\mathbf{r}_0}},  \label{bsix}
\end{eqnarray}
where we have used that for the vibrations parallel to the surface 
\begin{eqnarray}
\frac \partial {\partial x}D_{ll}(\mathbf{r},\mathbf{r}) &=&0, \\
\frac \partial {\partial x}D_{lk}(\mathbf{r},\mathbf{r}_0)\Biggr|_{\mathbf{r}%
=\mathbf{r}_0} &=&-\frac \partial {\partial x}D_{lk}(\mathbf{r}_0,\mathbf{r})%
\Biggr|_{\mathbf{r}=\mathbf{r}_0}=  \nonumber \\
&&\frac \partial {\partial x}D_{xz}(\mathbf{r},\mathbf{r}_0)\Biggr|_{\mathbf{%
r}=\mathbf{r}_0}\left( \delta _{xl}\delta _{zk}-\delta _{xk}\delta
_{zl}\right) .
\end{eqnarray}
For vibration normal to the surface only the diagonal elements of the
Green's function are non- vanishing and for this case (\ref{bthree}) reduce
to 
\begin{eqnarray}
\left( \mathbf{F}_{\mathit{fric}}\right) _z &=&u_04\pi A(\omega ,T)\omega
_0\int_{-\infty }^\infty \mathrm{d}\omega \frac \partial {\partial \omega
}\sum_{l=x,y,z}\Biggl\{\frac{\partial ^2}{\partial z\partial z^{\prime }}%
\Biggl[\mathrm{Im}D_{ll}(\mathbf{r},\mathbf{r}^{\prime },\omega )  \nonumber
\\
&&+\mathrm{Im}\Biggl(\frac{\alpha (\omega )D_{ll}(\mathbf{r},\mathbf{r}%
_0,\omega )D_{ll}(\mathbf{r}^{\prime },\mathbf{r}_0,\omega )}{1-\alpha
(\omega )D_{ll}(\mathbf{r}_0,\mathbf{r}_0,\omega )}\Biggr) \Biggr]\mathrm{Im}%
\frac{\alpha (\omega )}{1-\alpha (\omega )D_{ll}(\mathbf{r}_0,\mathbf{r}%
_0,\omega )}  \nonumber \\
&&+\Biggl(\frac \partial {\partial z}\mathrm{Im}\Biggl( \frac{\alpha (\omega
)D_{ll}(\mathbf{r},\mathbf{r}_0,\omega )}{1-\alpha (\omega )D_{ll}(\mathbf{r}%
_0,\mathbf{r}_0,\omega )}\Biggr)\Biggr)^2\Biggr\} _{\QATOP{\mathbf{r}=%
\mathbf{r}_0}{\mathbf{r}^{\prime }=\mathbf{r}_0}}
\end{eqnarray}

\thinspace

\end{document}